\documentclass[english,aps,prl,superscriptaddress,floatfix,notitlepage,reprint,show pacs]{revtex4-2}
\usepackage[utf8]{inputenc}
\usepackage[T1]{fontenc}
\usepackage{physics}
\usepackage{natbib}
\usepackage{amsthm}
\usepackage{amssymb}
\usepackage{dsfont}
\usepackage{amsmath}
\usepackage{bm}
\makeatletter

\usepackage{braket}
\usepackage{txfonts}
\usepackage{graphicx}
\usepackage{subcaption}
\usepackage{booktabs}

\usepackage[dvipsnames]{xcolor}
\usepackage[colorlinks=true,citecolor=blue,linkcolor=RubineRed,urlcolor=blue]{hyperref}
\usepackage{orcidlink}
\usepackage{url}
\usepackage{tikz}
\usepackage{rotating}

\begin{document}
\makeatletter
\def\@seccntformat#1{\csname the#1\endcsname.\quad}

\makeatother
\title{Finite temperature phase diagram of the extended Bose-Hubbard model in the presence of disorder}

\author{Madhumita Kabiraj \orcidlink{0009-0001-8509-9825}}
\email{madhumitakabiraj98@gmail.com}
\affiliation{Department of Physics, University of Calcutta, $92$ A.P.C. Road, Kolkata- $700009$}

\author{Raka Dasgupta \orcidlink{0000-0003-2148-4641}}
\email{rdphys@caluniv.ac.in}
\affiliation{Department of Physics, University of Calcutta, $92$ A.P.C. Road, Kolkata- $700009$}

\begin{abstract}
    We study the finite- and non-zero temperature phase diagram of the Extended Bose-Hubbard Model for both pure and disordered systems. Such a system can be experimentally realized by trapping ultracold Rydberg atoms in optical lattices. By regulating the Rydberg excitation level and the lattice spacing, the system can be engineered to effectively have (i) only the nearest-neighbor interaction and (ii) both nearest-neighbor and next-nearest-neighbor interactions. For both of these situations, we construct the mean-field phase diagrams. It is found that the presence of a non-zero temperature significantly changes the phase diagram because now there is a competition between quantum and thermal fluctuations. We observe that conventional Mott insulator (MI) or charge-density-wave (CDW) lobes vanish at higher temperatures. In a pure system, they melt into a normal fluid (NF). In contrast, the insulating phases that survive at high temperatures in the presence of disorder are the Bose glass and the normal fluid.  It is evident that the CDW lobes melt at a lower temperature and the Mott lobes melt at higher temperatures. These transition temperatures depend on the on-site and nearest-neighbor interaction strengths, respectively. It is also found that, with the addition of disorder, the insulating lobes are destroyed at a relatively lower temperature. The mathematical framework that we present here is capable of treating long-range interactions, disorder, and finite temperature simultaneously, and versatile enough so that it can be extended to study different forms of disorder or longer-range interactions. 
\end{abstract}

\maketitle

\section{Introduction}\label{sec:intro} In the last few decades, the Bose-Hubbard model (BHM) has remained one of the most widely acclaimed models in condensed matter physics. Although its initial conception was in the context of superconductivity, the Bose-Hubbard model later gained even more prominence with the advent of ultracold atom experiments and optical lattices. It represents the simplest Hamiltonian that describes strongly correlated ultracold bosons loaded in optical lattice potentials\cite{jaksch2005cold,morsch2006dynamics,bloch2008many,lewenstein2007ultracold}. This model has been extensively investigated using various theoretical techniques such as mean-field theory \cite{sheshadri1993superfluid,van2001quantum,amico1998dynamical}, exact diagonalization \cite{kashurnikov1996exact}, Quantum Monte Carlo (QMC) algorithm \cite{krauth1991mott,batrouni1992world}, strong coupling approach \cite{niemeyer1999strong,freericks1994phase,sengupta2005mott}, density matrix renormalization group (DMRG) \cite{kuhner1998phases,ranccon2011nonperturbative}, projection operator method \cite{trefzger2011nonequilibrium,dutta2012projection}, variational cluster approach \cite{koller2006variational,ejima2012characterization}, variational matrix product state technique \cite{kiely2022superfluidity}, etc. These studies successfully describe the quantum phase transition between the Mott-insulator (MI) and the superfluid (SF) phase. Several experimental works \cite{greiner2002quantum,bakr2010probing,sherson2010single} have demonstrated this transition by varying the depth of the lattice. 

Past studies on the Bose-Hubbard model mainly focus on zero-temperature systems, while a few ventured out to address its finite-temperature counterpart \cite{gerbier2007boson,trotzky2010suppression,fang2011quantum,mahmud2011finite}. In the non-zero temperature system, the phases are governed by thermal fluctuations as well, and the phase diagram changes appreciably. Both quantum and thermal fluctuations coexist in the presence of very low but non-zero temperature, leading to very interesting features. Such phase diagrams are of great physical relevance, as all realistic ultracold atom experiments belong to this regime. Some experiments have also been done on the BHM at finite temperature for $2D$ \cite{sherson2010single} (discussed about the melting of MI at higher temperature) and $3D$ \cite{nakamura2019experimental} (compared with theoretical predictions) optical lattices.

An interesting variant of the Bose-Hubbard model is the extended Bose-Hubbard model (EBHM), where long-range interactions are also present. Such a system can be designed with dipolar bosons \cite{baier2016extended} and Rydberg dressed atoms \cite{barbier2021extended,weckesser2024realization,li2018supersolidity}. Most of the theoretical works on EBHM have been done with nearest-neighbor (NN) interaction, using different techniques such as mean field approximation \cite{iskin2011route}, Gutzwiller variational method\cite{kovrizhin2005density,suthar2020staggered}, Quantum Monte-Carlo\cite{ohgoe2012ground},  DMRG \cite{kuhner2000one,pai2005superfluid,urba2006one,rossini2012phase}, etc. Experimentally, such a system has recently been realized using semiconductor dipolar excitons \cite{lagoin2022extended}. All of the above methods yield similar results and predict that in the phase diagram, two phases would arise in addition to MI and SF. One of them is a charge density wave (insulating phase with fractional occupancy), and another is the supersolid phase (a superfluid phase that has additionally broken continuous translational symmetry). 

Exciting new features appear in the Bose-Hubbard Model when some disorder is incorporated in the system. In experiments with ultracold atoms in optical lattices, such disorders have been imposed by several techniques, such as by a bichromatic lattice produced by two laser beams with incommensurate wavelengths \cite{deissler2010delocalization,fallani2007ultracold}, by adding another species in the system \cite{gavish2005matter,ospelkaus2006localization}, or alternatively, by using a speckle laser field \cite{fallani2007ultracold,billy2008direct,white2009strongly}. Disorder can be present in the BHM  in either the on-site potential, interparticle interaction\cite{gimperlein2005ultracold,niederle2015bosons} or tunneling\cite{bissbort2010stochastic,niederle2015bosons,sengupta2007quantum}. Bose glass (BG) phase, an insulator but compressible phase, emerges as an effect of disorder. 

In this work, we study the finite-temperature effects on the EBHM, something which is not well explored yet. We probe how the insulating lobes behave if the temperature is kept increasing, and how they gradually pave way for a normal fluid in both the absence and presence of disorder. We restrict ourselves to mean-field calculations only, because in the zero-temperature limit mean-field is proven to be a time-tested tool to bring out the key features of the model in a simple yet elegant manner. We expect that in the non-zero temperature regime, too, it would capture the essential results. If a more intricate analytical tool like the strong-coupling method \cite{niemeyer1999strong,freericks1994phase,sengupta2005mott} or the projection operator method \cite{trefzger2011nonequilibrium,dutta2012projection} is clubbed with the mean-field, the exact shapes of the lobes might change: but the overall qualitative features are expected to remain the same. 

The paper is organized as follows. The Hamiltonian of the EBHM containing disorder potential is presented in Sec. \ref{hamil}, and its possible implementation using Rydberg-dressed potential is discussed in Sec. \ref{sec:Ryd}. Then, in Sec. \ref{pure}, we construct the phase diagram of EBHM with nearest-neighbor interaction at non-zero temperatures (in the absence of disorder) using the mean-field perturbative method. The CDW and Mott lobes are found to vanish at higher temperatures, leading to normal fluid states. The transition temperatures are extracted in terms of the nearest-neighbor and on-site interaction parameters, respectively. In Sec. \ref{disorder}, the theory is further extended to incorporate the effects of a uniform disorder in the on-site potential, and the fate of all insulating states (including Bose glass) is studied at higher temperatures.  In Sec. \ref{long-range}, the range of long-range interaction is increased up to the next-nearest neighbors (NNN), and the modified phase diagrams are studied. We present a summary and outline possible future directions in Sec. \ref{con}.

\section{The Model}\label{sec:model}
\subsection{Extended Bose-Hubbard Hamiltonian}\label{hamil}
We consider a bosonic system in an optical lattice described by the Extended Bose-Hubbard Hamiltonian,
	\begin{equation}
 \label{eq:1}
 \begin{split}
		H=-t\sum_{\langle{ij}\rangle}(b_i^\dagger b_j+b_j^{\dagger}&b_i)-\sum_{i}(\mu+\epsilon_{i})n_i\\&+\frac{U}{2}\sum_{i}n_i(n_i-1)+V\sum_{\langle{ij}\rangle}n_in_j
	 \end{split}
	\end{equation}
where $t$ is the hopping amplitude (here we consider hopping between nearest neighbors only), $b_i$ and $b_i^\dagger$ are bosonic annihilation and creation operators respectively, $n_i$ is the number operator, $U$ is the on-site interaction strength between bosons, $V$ is the strength of nearest neighbor interaction, and $\mu$ is the chemical potential. The disorder is expressed as an additional on-site energy $\epsilon_{i}$ at site $i$. We study both pure ($\epsilon_{i}=0$) and disordered ($\epsilon_{i}\neq 0$) situations at non-zero temperatures.
\begin{figure*}[t!]
    \centering
    \begin{subfigure}[b]{0.4\linewidth}
        \centering
        \includegraphics[width=0.8\textwidth]{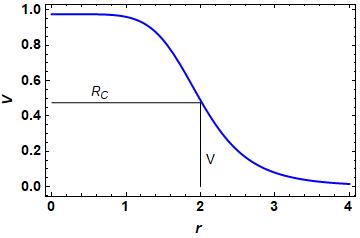}
        \caption{Rydberg potential as a function of lattice interatomic separation. Here we take $R_C=a$ (here $2 \mu{m}$) for the NN interaction where the potential is almost half of the on-site interaction.}
        \label{ryd1}
    \end{subfigure}
    \begin{subfigure}[b]{0.4\linewidth}
        \centering
        \includegraphics[width=0.8\textwidth]{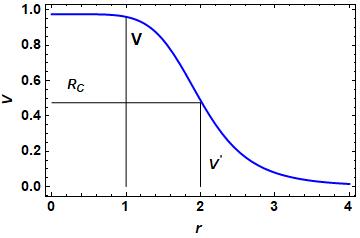}
        \caption{Potentials considering both the nearest and next-nearest neighbor interactions. Lattice spacing is $1 \mu{m}$. Here $V=0.96U$ and $V^\prime=0.48U$.}
        \label{ryd2}
    \end{subfigure}
    \caption{Rydberg potentials}
    \label{fig:1}
\end{figure*}
\subsection{Experimental realization of long-range interactions }\label{sec:Ryd}
In many-body physics, Rydberg-dressed atoms in optical lattice potentials can be excellent tools to engineer long-range interactions. Such a system can be easily adapted to be used in EBHM with only NN interactions\cite{weckesser2024realization}. For Rydberg systems, the long-range, soft-core potential between two atoms in the ground state \cite{chougale2016ab} is,
\begin{equation}
\label{Vij}
    V_{ij}=\left(\frac{\Omega}{2{|\delta|}}\right)^4 \frac{C_6}{R_c^6+r_{ij}^6}
\end{equation}
Here  $R_c$ is the soft-core radius (that determines the range of interaction), and $r_{ij}$ is the separation between the $i^{th}$ and $j^{th}$ atoms, $C_6$ is the interaction coefficient, $\Omega$ is the Rabi frequency, and $\delta$ is the detuning. One can regulate the range of interaction ($R_c$) by controlling the detuning $\delta$ and $C_6$ (that depends on the Rydberg excitation level), through the relation $R_c=[C_6/2\hbar{|\delta|}]^{\frac{1}{6}}$.
To illustrate the nature of the long-range interactions, we present the $V$ vs. $r$ plots (Fig.\ref {fig:1}) for two sets of realistic experimental parameters taken from \cite{chougale2016ab} for $^{87}Rb$ atoms.

In Fig.\ref{ryd1} the parameters are lattice spacing $a=2\,{\mu}m$, $\Omega/\delta=0.1$, $\delta=100\,MHz$, $C_6=10^6\, MHz\, {\mu}m^6$, and $U=100\,Hz$, resulting in $R_c\,=\,a$. So, effectively, it can be treated as a lattice with constant NN interaction, and the longer-range interactions can be neglected altogether. From Eq.\eqref{Vij}, we get the NN strength $V=0.48U$ (as shown in Fig.\ref{ryd1}). 
 
In Fig.\ref{ryd2} all parameters are the same as in Fig.\ref{ryd1} except for the lattice spacing that has been changed such that $R_c\,=\,2a^\prime$ ($a^\prime$ is the new lattice spacing, here it is $1\,{\mu}m$). In this case, both NN and NNN interactions are important, and all subsequent longer-range interactions become negligibly small. Inserting in Eq.\eqref{Vij}, we find that the NN strength $V\,=\,0.96U$, NNN strength $V^\prime\, =\,0.48U$, as evident from Fig. \ref{ryd2}.
  
All these values of $V$ and $V^\prime$ have been used in the subsequent sections.

\section{Phase diagram of Pure System}\label{pure}
As in ref. \cite{iskin2011route}, we split the entire lattice into two sublattices (say A and B) so that the nearest-neighbor sites belong to a different sublattice.
We use the following decoupling of hopping term from  \cite{iskin2011route} and \cite{kurdestany2012inhomogeneous},
\begin{align*}
	b_i^{\dagger}b_j & \rightarrow {\langle{b_i^{\dagger}}\rangle}b_j\,+b_i\langle{b_j^{\dagger}}\rangle\,-\langle{b_i^{\dagger}}\rangle{\langle{b_j}\rangle}\\
	& = 	b_i^{\dagger}\psi\,+\psi^*b_i\,-|\psi|^2
\end{align*}
where $\psi = \langle{b}\rangle$ is the superfluid order parameter.

The Hamiltonian is thus reduced to a single-site mean-field Hamiltonian and can be written as follows,
\begin{multline}\label{eq:2}
	{H} = -t(\phi_ib_i^{\dagger}\,+\,\phi_i^*b_i) + {\frac{t}{2}}(\psi_i^*\phi_i\,+\,\psi_i\phi_i^*) \\ +\frac{U}{2}n_i(n_i-1) + zVn_i\bar{n_i} - {\mu}n_i
\end{multline}
where $\psi \equiv \sum\limits_{j}\psi_j$ is the sum of the order parameters at sites j neighboring to site i and $\bar{n_i} \equiv \sum\limits_{j}n_j$.

Now, for a bipartite lattice, each site on sublattice \textbf{A(B)} has z nearest neighbors, each of which belongs to sublattice \textbf{B(A)}.
Therefore, $\psi_i = \psi_A, \phi_i = z\psi_B$ and $\bar{n_i} = zn_B\,$ for $i \in A$ and
$\psi_i = \psi_B, \phi_i = z\psi_A$ and $\bar{n_i} = zn_A\,$ ,  for $i \in B$. 
In \cite{iskin2011route}, using mean-field decoupling considering the hopping as a perturbation the phase  equation has been obtained.
\begin{equation}
\begin{split}\label{eq:3}
	\frac{1}{z^2t^2} = & \left[\frac{n_{A} + 1}{Un_{A} + zVn_{B} -\mu} - \frac{n_{A}}{U(n_{A} - 1) + zVn_{B} - \mu}\right] \\& \times \left[\frac{n_{B} + 1}{Un_{B} + zVn_{A} -\mu} - \frac{n_{B}}{U(n_{B} - 1) + zVn_{A} - \mu}\right]
\end{split}
\end{equation}
Here, the two terms come from sublattices A and B, respectively.
Plots of Eq. \eqref {eq:3} for different sets of values of $n_A$ and $n_B$ give the zero-temperature phase diagram showing alternate charged density wave (CDW) and MI (mean-field \cite{iskin2011route}, strong-coupling perturbation \cite{iskin2009strong}). The phase diagram is almost the same as the zero-temperature diagram obtained by QMC or Gutzwiller variational method \cite{kovrizhin2005density,iskin2011route,ohgoe2012ground}. However, these other methods also predict another compressible phase known as supersolid (SS) in the region with non-zero $\psi$ along with the traditional superfluid phase,  a phase that cannot be constructed using the mean-field analytical approach or by the strong-coupling perturbation theory \cite{iskin2009strong}.  In our present work, the focus is on the insulating phases primarily, so we restrict ourselves to the mean-field treatment, and the phase boundary between SF and SS is not studied.

Now, to include temperature effects in this system, we take the thermal average \cite{gerbier2007boson} in Eq.\eqref{eq:3}. 
Since the two terms in \eqref{eq:3} come from two different sublattices $A$ and $B$, respectively, the averages are taken separately.

The partition function for the unperturbed system (i.e., when hopping $t=0$) is given by
\begin{equation}\label{eq:6}
	z = \sum_{n_A = 0}^{\infty}\sum_{n_B = 0}^{\infty} e^{-\beta {E_{n_A,n_B}}}
\end{equation}
Where $E_{n_A, n_B}= (U/{2})n_A(n_A-1) +(U/{2})n_B(n_B -1) +zVn_An_B -\mu(n_A +n_B)$ is the system's total unperturbed energy.
Therefore, the thermal average of $(1/{zt})$ for two sublattices will be,
\begin{equation}
	\label{eq:7A}
	\left<\frac{1}{zt}\right>_A = \frac{\sum_{n_A = 0}^{\infty}\sum_{n_B = 0}^{\infty} e^{-\beta {E_{n_A,n_B}}}\times{f_A\{n_A , n_B\}}}{\sum_{n_A = 0}^{\infty}\sum_{n_B = 0}^{\infty} e^{-\beta {E_{n_A,n_B}}}}
\end{equation} and
\begin{equation}
	\label{eq:7B}
	\left<\frac{1}{zt}\right>_B = \frac{\sum_{n_A = 0}^{\infty}\sum_{n_B = 0}^{\infty} e^{-\beta {E_{n_A,n_B}}}\times{f_B\{n_A , n_B\}}}{\sum_{n_A = 0}^{\infty}\sum_{n_B = 0}^{\infty} e^{-\beta {E_{n_A,n_B}}}}
\end{equation}
where
	\begin{equation*}
		\label{eq:8}
	\begin{split}
	&{f_A\{n_A , n_B\}}
 \\&=\left(\frac{n_{A} + 1}{Un_{A} + zVn_{B} -\mu} - \frac{n_{A}}{U(n_{A} - 1) + zVn_{B} - \mu}\right)  
	\end{split}
	\end{equation*}and
    \begin{equation*}
        \begin{split}
           &{f_B\{n_A , n_B\}} \\&=\left(\frac{n_{B} + 1}{Un_{B} + zVn_{A} -\mu} - \frac{n_{B}}{U(n_{B} - 1) + zVn_{A} - \mu}\right)
        \end{split}
    \end{equation*}

\begin{figure}
        \centering
        \includegraphics[width=0.8\linewidth]{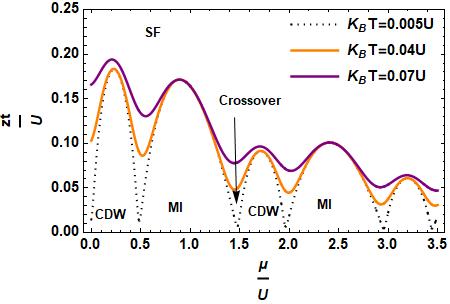}
        \caption{Phase diagram of extended Bose-Hubbard model at various temperatures with NN interaction strength $V=0.48U$ (obtained from Rydberg interaction in sec. \ref{sec:Ryd}).}
        \label{fig:2}
    \end{figure}

	Equations \eqref{eq:7A} and \eqref{eq:7B} are combined together and plotted as ($zt/U$ vs. $\mu/U$) at different temperatures keeping the NN interaction strength $V=0.48U$ that is obtained from Rydberg potential (done in Sec.\ref{sec:Ryd}) (Fig.\ref{fig:2}). In these equations, if we put $V=0$ in the zero temperature limit, they produce the well-known phase diagram of BHM. At very low temperatures ($K_BT=0.005U$), the diagram is identical to the zero-temperature phase diagram, i.e. alternate CDW and MI phases appear (dotted curves in Fig.\ref{fig:2}). The width of each lobe depends on the interaction strengths (both $U$ and $V$). As the temperature increases, the superfluid phase shifts towards higher values of hopping $t$ (orange curves). The superfluid phase is a superposition state with strong quantum fluctuations in the number of atoms per site in an optical lattice. At higher temperatures, thermal fluctuation dominates over quantum fluctuation resulting in a shift of SF, and the SF region near the edges is replaced by a smooth cross-over region between the MI and CDW phases. This phase has finite compressibility and is named a normal phase in \cite{gerbier2007boson}. Fig.\ref{fig:2} indicates the boundary between the SF and insulator-like phases (MI and CDW). For a better understanding of the boundaries between different insulating phases in the crossover, we plot the number density and compressibility at zero hopping. At a temperature $T$, the expectation value of the boson density of the system becomes,
    \begin{equation}
         \langle{\rho}\rangle = \frac{\sum_{n_A = 0}^{\infty}\sum_{n_B = 0}^{\infty} e^{-\beta {E_{n_A,n_B}}}\times{(n_A+n_B)/2}}{\sum_{n_A = 0}^{\infty}\sum_{n_B = 0}^{\infty} e^{-\beta {E_{n_A,n_B}}}}
         \label{den}
    \end{equation}
     and  the compressibility (the variation of number density with respect to chemical potential) is,
    \begin{equation}
        \kappa = \frac{\partial{\langle{\rho}\rangle}}{\partial\mu}
        \label{com}
    \end{equation}
\begin{figure}
    \centering
    \includegraphics[width=0.9\linewidth]{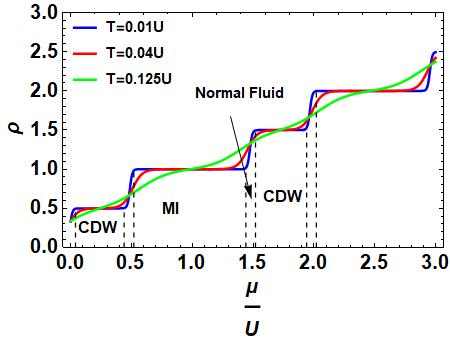}
    \caption{Density plot for NN interaction strength $V=0.48U$ at different temperatures. Black dashed lines separate MI, CDW and normal fluid at temperature $K_BT=0.01U$ (blue curve).}
    \label{fig:3}
\end{figure}
    
\begin{figure}
        \centering
        \includegraphics[width=0.8\linewidth]{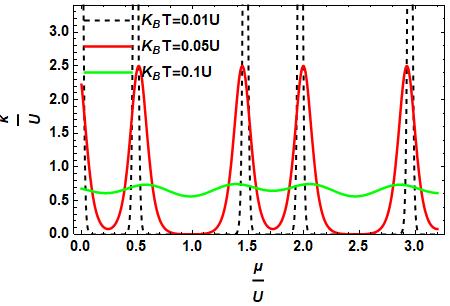}
        \caption{Compressibility plot for insulating phases.}
        \label{fig:4}
    \end{figure}

    At sufficiently low temperatures, the compressibility curve shows sharp peaks at the boundary (black dashed lines in Fig.\ref{fig:4}), indicating the separation between CDW and MI. With an increasing temperature, the width of the compressible region (normal fluid) increases. Consequently, the MI and CDW regions gradually decrease, and above certain transition temperatures, they completely disappear. This transition or melting temperature depends on the width of the insulating lobes. If all energies are scaled in units of on-site interaction $U$, the width of the MI lobes is fixed to unity at zero temperature. Thus, the melting temperature of the Mott lobes is the same for all values of interaction $V$. The transition temperature for the MI is approximately $K_BT^*=0.1U$. Since $V$ is less than $U$, CDW disappears at lower temperatures. It survives up to $K_BT^*\thickapprox0.08V$. Above the transition temperature of the MI, the entire system becomes a normal compressible fluid. In the density vs. chemical potential plot (Fig.\ref{fig:3}) we draw a series of black dashed lines that separate the incompressible phases (MI, CDW) and the compressible ones (normal fluid) at a temperature $K_BT=0.01U$ (blue line). A sufficiently low value of  $\kappa$  can be used to define the incompressible phase \cite{fang2011quantum,mahmud2011finite} (for that, we choose $\kappa\leq 0.01U$) as the number density remains almost constant in this range.   
 
	\section{Phase diagram for Disordered system}\label{disorder}
    In most of the previous works, the disorder is introduced mainly in the BHM through the onsite potential term of the Hamiltonian (Eq.\eqref{eq:1}), which is essentially a diagonal disorder. The potential $\epsilon_i$ can obey different probability distributions such as uniform box distribution \cite{fisher1989boson,bissbort2009stochastic,krutitsky2006mean,bissbort2010stochastic,soyler2011phase,gurarie2009phase}, Gaussian form \cite{wu2008minimal,de2018properties}, speckle \cite{kruger2009anomalous,bissbort2010stochastic}, independent random variables following a probability distribution \cite{buonsante2009gutzwiller,pollet2013review}, and other similar forms
\cite{freericks1996strong,sheshadri1995percolation,sengupta2007quantum,wang2016bogoliubov}. 
  Of these, we choose the uniform form of disorder. The local disorder $\epsilon_i$ in Hamiltonian \eqref{eq:1} is distributed uniformly in the interval $\left[-{\Delta}/{2},{\Delta}/{2}\right]$ \cite{krutitsky2006mean} with,
\begin{equation}\label{eq:9}
	p(\epsilon) = \frac{1}{\Delta}\left[\Theta\left(\epsilon+\frac{\Delta}{2}\right)-\Theta\left(\epsilon-\frac{\Delta}{2}\right)\right]
\end{equation}
	where $\Theta$ is the Heaviside step function. We use the analytical mean-field approach (discussed in Appendix \ref{A}) for EBHM in the presence of disorder. To deal with any disorder following a probability distribution, we need to do a disorder average of the form,
    \begin{equation}
        \overline{F} = \int_{-\infty}^{\infty}F\,p(\epsilon)\,d\epsilon
        \label{eq:10}
    \end{equation}
    or,\begin{equation*}
        \overline{F} = \int_{\mu-\infty}^{\mu+\infty}F\,p(\Tilde{\mu}-{\mu})\,d\Tilde{\mu}
    \end{equation*}
    where, $\Tilde{\mu}\,=\,\mu+\epsilon$. In our system ${F}$ is the quantity $1/zt$ in equations \eqref{eq:7A} and \eqref{eq:7B}. Therefore, both the thermal-averaged and disorder-averaged expression of $1/zt$ is
    \begin{equation}
        \begin{split}
            &\overline{\left<\frac{1}{zt}\right>}_A \\&= \int_{\mu-\frac{\Delta}{2}}^{\mu+\frac{\Delta}{2}}\frac{\sum_{n_A = 0}^{\infty}\sum_{n_B = 0}^{\infty} e^{-\beta {E_{n_A,n_B}}}\times{f_A\{n_A , n_B\}}}{\sum_{n_A = 0}^{\infty}\sum_{n_B = 0}^{\infty} e^{-\beta {E_{n_A,n_B}}}}p(\Tilde{\mu}-{\mu})\,d\Tilde{\mu}
        \end{split}
        \label{dis1}
    \end{equation}
    for sublattice A, and similarly for sublattice B. 
    As the uniform disorder $\epsilon$ is bounded between $(-\Delta/2)$ and $(\Delta/2)$, the integration is from $\left(\mu-(\Delta/2)\right)$ to ${\left(\mu+(\Delta/2)\right)}$. For the entire system, the equation for the boundary of the insulating-non-insulating phases becomes \begin{equation}
        \overline{\left<\frac{1}{zt}\right>}=\sqrt{\overline{\left<\frac{1}{zt}\right>}_A\times\overline{\left<\frac{1}{zt}\right>}_B}
        \label{dis2}
    \end{equation}

    These phase boundaries are plotted for different values of NN interaction strengths, disorder strengths, and temperatures, and presented in the next two subsections. There is a competition between the strengths of the disorder and the NN interaction, and the phases are affected by this.  
    
    \begin{figure*}[t!]
    \centering
    \begin{subfigure}[b]{0.4\linewidth}
        \centering
        \includegraphics[width=1.0\textwidth]{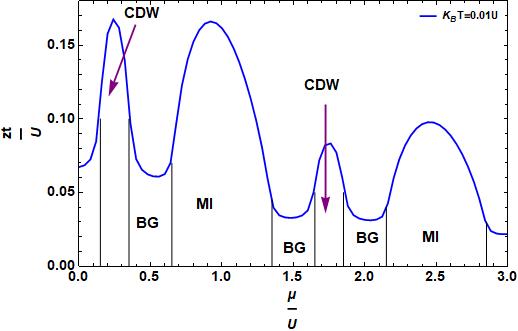}
        \caption{Phase diagram for EBHM with  $V=0.48U$, $\Delta=0.3U$, $K_BT=0.01U$. Widths of CDW and MI phases shrink and an insulating but compressible BG appears between them.}
        \label{5a}
    \end{subfigure}
    \begin{subfigure}[b]{0.4\linewidth}
        \centering
        \includegraphics[width=1.0\textwidth]{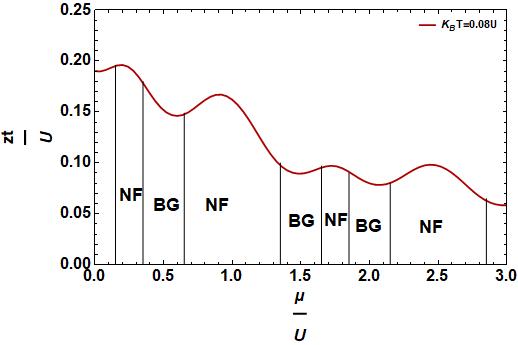}
        \caption{At higher temperature $K_BT=0.1U$. MI and CDW disappear, NF and Boss-glass phases exist within the insulating boundary.}
        \label{5b}
    \end{subfigure}
    \caption{Phases at different temperatures for $V>\Delta$.}
    \label{fig:5}
\end{figure*}
\subsection{Phase Diagrams}
\subsubsection{Phase diagram for \texorpdfstring{$V>\Delta$}{V>Delta}}
  First, we consider that the NN interaction term is greater than the disorder strength $\Delta$ and plot the phase diagram (Fig.\ref{fig:5}). The interaction potential $V$ is obtained from the Rydberg dressed interaction (done in Sec.\ref{sec:Ryd}). The phase diagram thus obtained has signatures of the effects of disorder, NN interaction, and thermal fluctuations all at once. The influences of each term are listed below.
\begin{itemize}
    \item NN interaction $\rightarrow$ Appearance of the CDW phase (as seen in Sec. \ref{pure}).
    
    \item Disorder $\rightarrow$ If $\Delta$ is less than onsite interaction $U$, gaps of width $\Delta$ arise between incompressible lobes by squeezing them, which represent compressible fluids (Fig.\ref{5a}). This compressible insulating phase is the Bose glass phase. This was observed by the mean-field perturbative method in \cite{krutitsky2006mean} and by Green's function approach in \cite{souza2021green} for the BHM with uniform disorder. If the disorder $\Delta$ is greater than or equal to $U$, MI and CDW lobes will vanish, and only Bose glass would exist at all temperatures.
    
    \item Temperature $\rightarrow$ The effect of non-zero temperature on the MI and CDW is already discussed in Sec.\ref{pure}. At zero temperature, the Bose glass and superfluid boundary lies on the $\mu/U$ axis \cite{gao2017phase}. As temperature increases, the boundary shifts towards higher values of hopping (Fig.\ref{5a}). For BG, the effect of temperature is the expansion in the higher hopping region, with the width fixed at $\Delta$. Although for MI and CDW the finite temperature effect is the same as seen in Sec.\ref{pure}, i.e., near the edges of CDW and MI, the normal fluid appears due to thermal fluctuations, where the compressibility becomes finite. After a certain temperature, depending on $\Delta$, they completely melt into the normal fluid. Therefore, CDW and MI disappear one after another, and only BG and NF are present on the insulating side (Fig.\ref{5b}). This is discussed on the basis of compressibility and IPR in the following subsection.
\end{itemize}

\subsubsection{Phase diagram for \texorpdfstring{$V<\Delta$}{V<Delta}}
If the disorder strength $\Delta$ is larger than the NN interaction $V$, the disorder effect blocks the emergence of CDW phases. There is no contribution of long-range interaction till $V\leq\Delta$. In the insulating region, the phase diagram contains only MI and BG phases (Fig.\ref{fig:7}) at very low temperatures and NF between BG, MI crossover at moderate temperatures until MI melts.
 \subsection{Classifying The Phases}\label{clas}

 We categorize different insulating phases on the basis of two quantities: $i)$ compressibility, $ii)$ inverse participation ratio.
\subsubsection{Compressibility:}
The phase diagram (Fig.\ref{fig:5}) is obtained with respect to the order parameter $\psi$, which only indicates the boundary between the insulator and the superfluid. Thus, the boundaries of the MI, CDW from the normal fluid cannot be directly extracted. Therefore, we calculate a disordered compressibility average to determine the boundary between the insulating phases in the crossover region. A detailed derivation of the expression of compressibility is done in Appendix \ref{B}. The Compressibility of the insulating phases at zero hopping is,
\begin{equation}\label{compressibility}
\begin{split}
    \kappa=&\frac{1}{2\Delta}\left[{\frac{\sum_{n_A=0}^\infty\sum_{n_B=0}^\infty (n_A+n_B) e^{-\beta E_{n_A,n_B}\left(\mu+\frac{\Delta}{2}\right)}}{\sum_{n_A=0}^\infty\sum_{n_B=0}^\infty  e^{-\beta E_{n_A,n_B}\left(\mu+\frac{\Delta}{2}\right)}}}\right.\\
    &\left.-{\frac{\sum_{n_A=0}^\infty\sum_{n_B=0}^\infty (n_A+n_B) e^{-\beta E_{n_A,n_B}\left(\mu-\frac{\Delta}{2}\right)}}{\sum_{n_A=0}^\infty\sum_{n_B=0}^\infty  e^{-\beta E_{n_A,n_B}\left(\mu-\frac{\Delta}{2}\right)}}}\right]
    \end{split}
\end{equation}
\begin{figure}
    \centering
    \includegraphics[width=1.0\linewidth]{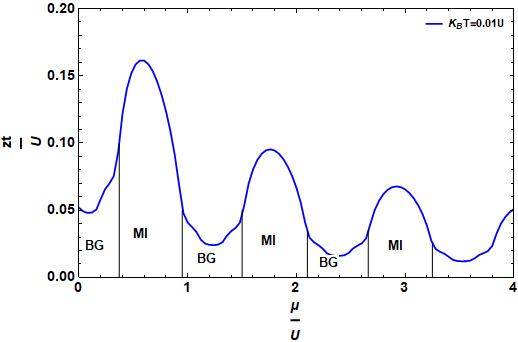}
    \caption{Phase diagram for $V<\Delta$. Here $\Delta=0.4U$, $V=0.15U$ and $K_BT=0.01U$. No effect of $V$ since it is less than $\Delta$.}
    \label{fig:7}
\end{figure}
\begin{figure}
    \centering
    \includegraphics[width=1.0\linewidth]{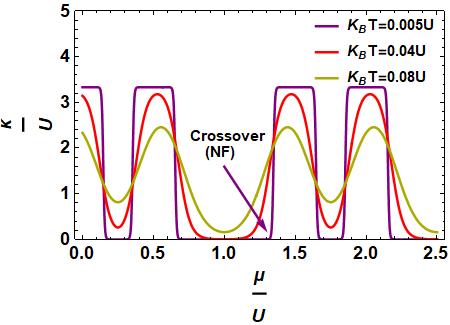}
    \caption{Compressibility at different temperatures with a fixed disorder strength $\Delta=0.3U$. At $K_BT=0.005U$ (the purple curve), $\kappa$ is zero for MI and CDW; and the non-zero blocks are the Boss glass. At $K_BT=0.04U$ (red curve), CDW is completely replaced by NF and at $K_BT=0.08U$ (green curve), MI also vanishes.}
    \label{fig:6}
\end{figure}

Fig.\ref{fig:6} shows the compressibility for different temperatures with $\Delta=0.3U$ and $V=0.5U$. At sufficiently low temperature (at $K_BT=0.005U$, purple curves in Fig.\ref{fig:6}) compressibility is finite in the Bose glass region and almost zero in the MI and CDW region. As temperature increases, the widths of the compressibility curve spread from the boundary towards the incompressible lobes (red curves in Fig.\ref{fig:6}). This crossover region is the normal fluid phase. Therefore, in the intermediate temperature range, along with CDW and MI, BG is present due to disorder, and NF is present due to thermal fluctuation. After a certain increase in temperature, depending on disorder and interaction strengths CDW and MI completely melt into the normal compressible fluid (green curve in Fig.\ref{fig:6} at $K_BT=0.08U$). Therefore, at high temperatures, BG and NF exist as insulators. 
In the compressibility plot (Fig.\ref{fig:6}), we take $\kappa\leq0.01U$ as the insulator limit.

\begin{figure}
    \centering
    \includegraphics[width=0.9\linewidth]{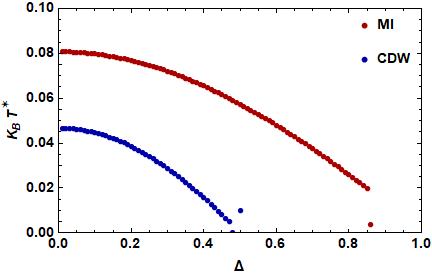}
    \caption{Transition temperature ($K_BT^*$) vs. $\Delta$ plot for CDW (blue) and MI (red) done at $U=1$ and $V=0.5$.}
    \label{transitionT}
\end{figure}

 The transition temperatures ($T^*$) up to which the incompressible phases survive depend on all interaction strengths as well as disorder. In Fig.\ref{transitionT} the variation of transition temperatures with disorder is shown for the CDW (blue line) and MI (red line) with $U$ and $V$ fixed at $1$ and $0.5$ respectively. Since $U$ is larger than $V$, $T^*$'s are larger for MI. For both phases, $T^*$ decreases with $\Delta$ because with increasing $\Delta$ the width of incompressible lobes gradually shrinks. So they melt at lower temperatures for large $\Delta$. For $V=0.5$, the transition temperature comes close to zero at $\Delta=0.5$ for the CDW lobes. This is why CDW is not present for $V<\Delta$ in Fig.\ref{fig:7}.

\begin{figure}
    \centering
    \includegraphics[width=1.0\linewidth]{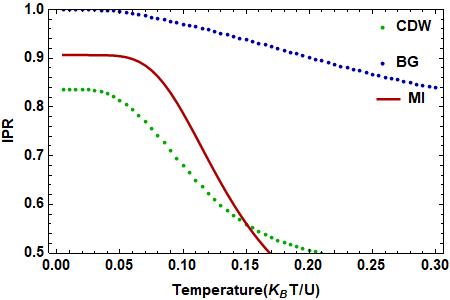}
    \caption{Variation of IPR with temperature. Here, $U=1$, $V=0.5$ and $\Delta=0.3$. The fall in the MI and CDW indicates the transition into NF, while the flat blue curve of IPR reflects that the BG is almost unaffected.}
    \label{ipr}
\end{figure}
 \subsubsection{Inverse participation ratio}\label{IPR}

 Another physical quantity that can be used for the classification of different insulators at finite temperature is the inverse participation ratio (IPR). It is a measure of localization \cite{perera2015energy,mansikkamaki2021phases}. For a single particle eigenstate $\Psi_i$, IPR is expressed as, $IPR=\sum_i|\Psi_i|^4/[\sum_i|\Psi_i|^2]^2$. In the phase space, IPR is one if the state is localized and close to zero for completely delocalized state. In our system among the insulating phases MI, CDW and BG have localized wavefunctions while normal fluid is a delocalized phase. Hence temperature average of IPR (thermal IPR) can justify the melting. We use exact diagonalization for this purpose. We take a system size $L=7$, and maximum number of bosons $N_{max}=7$. Fig.\ref{ipr} is a plot of the temperature dependence of thermal IPR for MI, CDW and BG at $V=0.5U$ and disorder strength  $\Delta=0.3U$ with disorder average over $100$ realizations. The chemical potential values are taken at the centre of each lobe. Fig.\ref{ipr} shows that for the MI (red curve) and CDW (green dotted curve) phases, IPR remains high at low temperature and falls off at a certain temperature. This indicates the transition into the delocalized normal fluid. But here the transition temperature is higher than that we get from Fig.\ref{transitionT} due to finite system size. For the BG phase, thermal IPR is high ($0.85-1$) at high temperatures, which signifies the existence of BG along with NF at higher temperatures. 
\begin{table}[h]
\centering
\caption{Classification of different insulating phases.}
 \begin{tabular}{c@{\hspace{1cm}}c@{\hspace{0.5cm}} c@{\hspace{0.5cm}}c}
    \hline
    \hline
     \textbf{Phase} & {$\mathbf{\psi}$} & {$\mathbf{\kappa}$} & \textbf{IPR} \\
     \hline
     MI and CDW & Zero & Zero & High\\
     \vspace{0.1cm}
     BG & Zero & Finite & High\\
     \vspace{0.1cm}
     NF & Zero & Finite & Low to medium\\
     \hline
     \hline
 \end{tabular}
 \end{table}

\section{When Next-nearest Neighbor interaction is also present in the system}\label{long-range}
\subsection{Pure System}\label{NNI+p}
\begin{figure}
    \centering
    \includegraphics[width=1.0\linewidth]{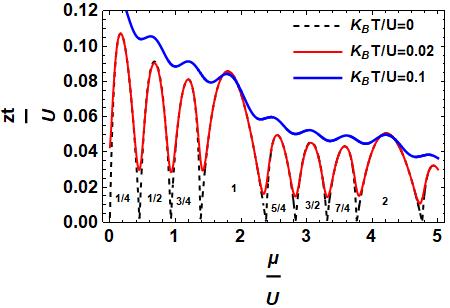}
    \caption{Phase diagram for $V/U=0.9$, $V^\prime/U=0.48$ (calculated from the Rydberg soft-core potential (Fig.\ref{ryd2} in Sec.\ref{sec:Ryd})). The numbers inside the lobes are the average boson density for that particular lobe.}
    \label{fig:8}
\end{figure}

We now consider the interaction to be present up to the next-nearest neighbor (NNN). This is easily achievable through the Rydberg atoms as illustrated in Fig.\ref{ryd2} in Sec.\ref{sec:Ryd}. The Hamiltonian for this system would be,
\begin{equation}
 \label{nnn1}
 \begin{split}
		{H}=&-t\sum_{\langle{ij}\rangle}(b_i^\dagger b_j+b_j^{\dagger}b_i)-\sum_{i}(\mu+\epsilon_{i})n_i\\&+\frac{U}{2}\sum_{i}n_i(n_i-1)+V\sum_{\langle{ij}\rangle}n_in_{j}+V^\prime\sum_{\langle{ik}\rangle}n_in_{k}
	 \end{split}
	\end{equation}
    here $V$ and $V^\prime$ are the repulsive interactions between the bosons at NN and NNN sites, respectively. 
    We can write the mean-field Hamiltonian for this system as follows,

   \begin{multline}\label{eq:MF}
	{H}^{MF} = -t(b_i^{\dagger}\phi_i\,+\,\phi_i^*b_i) + {\frac{t}{2}}(\psi_i^*\phi_i\,+\,\psi_i\phi_i^*) \\ +\frac{U}{2}n_i(n_i-1) + Vn_i\bar{n_i}+V^\prime\,n_i\bar{n_i^\prime} - {\mu}n_i
\end{multline}
Here $\bar{n_i}=\sum_j n_j$ (comes from NN sites) and $\bar{n_i^\prime}=\sum_k n_k$ (from NNN sites). To take account of the NNN term, we take a lattice of the form \textbf{ABCDABCDAB...}. So, for one sublattice, suppose for sublattice \textbf{A}, sublattices \textbf{B} and \textbf{D} are nearest neighbor sites ($z/{2}$ NN of each sublattice), and z no. of next-nearest neighbors belong to sublattice \textbf{C}. Therefore, for $i\in{A}$, $\psi_i=\psi_A$, $\phi_i=({z}/{2})(\psi_B+\psi_D)$, $\Bar{n_i}=({z}/{2})(n_B+n_D)$, and $\Bar{n_i^\prime}=zn_C$. This leads to the following mean-field Hamiltonian for sublattice \textbf{A},
    \begin{multline}
        {H}_A^{MF}=-\frac{zt}{2}(\psi_B+\psi_D)(b_A+b_A^\dagger)+zt\psi_A(\psi_B+\psi_D)\\
        +\frac{U}{2}n_A(n_A-1)+\frac{zV}{2}n_A(n_B+n_D)+zV^\prime{n_A}n_C-\mu{n_A}
    \end{multline}
    and similar expressions can be found for other sublattices. Total Hamiltonian of the system is a sum of mean-field Hamiltonians for all the sublattices, i.e, \begin{equation}
        {H}^{MF}={H}_A^{MF}+{H}_B^{MF}+{H}_C^{MF}+{H}_D^{MF} 
    \end{equation}
Here we also treat the hopping term as a perturbation and calculate order parameters (see Appendix \ref{A}). The correction to the ground state energy leads to the following four coupled equations, which are derived in Appendix \ref{A} (equations \eqref{A6} and \eqref{A8}),
\begin{equation}
\begin{split}
    &\psi_A=\frac{zt}{2}\left(\psi_B+\psi_D\right)\times{f_A}\\
    &\psi_B=\frac{zt}{2}\left(\psi_A+\psi_C\right)\times{f_B}\\
    &\psi_C=\frac{zt}{2}\left(\psi_B+\psi_D\right)\times{f_C}\\
    &\psi_D=\frac{zt}{2}\left(\psi_A+\psi_C\right)\times{f_D}
    \end{split}
\end{equation}
where $f_A$, $f_B$, $f_C$, $f_D$ are functions of interaction potentials and chemical potential written in Appendix \ref{A}.
Eliminating all $\psi's$ from the above equations we get,
\begin{equation}
    \frac{1}{z^2t^2}=(f_A+f_D)\times(f_B+f_C)
\end{equation}. This equation gives a phase diagram at zero-temperature (black dashed line in Fig.\ref{fig:8}). In the phase diagram, three types of incompressible phases arise due to the contribution of three different interactions. They are
\begin{itemize}
    \item MI  $\rightarrow$ appears from on-site interaction, has integer boson-density.
    \item CDW $1$ $\rightarrow$ effect of NN interaction, particle density is $1/2$, $3/2$, $5/2$ and so on.
    \item CDW $2$ $\rightarrow$ as a result of NNN interaction, the MI and CDW phases are surrounded by a new incompressible phase having the same number of particles per site in three sublattices and different in any one sublattice (e.g., for the first lobe, it is $0$ for \textbf{A}, \textbf{B} and \textbf{C} and $1$ for \textbf{D}. So. the particle density is ${1}/{4}$, ${3}/{4}$, ${5}/{4}$, so on. Since this phase also has fractional density, we call it CDW $2$. This is also illustrated in Table \ref{tab 1}.
\end{itemize}

\begin{table}[h]
    \centering
    \begin{tabular}{c|c|c|c|c}
    \toprule
        Phase & A & B & C & D \\
    \midrule    
         CDW 2 & 1 & 0 & 0 & 0  \\
         CDW 1 & 1 & 0 & 1 & 0  \\
         CDW 2 & 1 & 1 & 1 & 0  \\
         MI & 1 & 1 & 1 & 1  \\
     \bottomrule    
    \end{tabular}
    \caption{Number of bosons in different sublattices indicating different insulating phases.}
    \label{tab 1}
\end{table}

\begin{figure}
    \centering
    \includegraphics[width=1.0\linewidth]{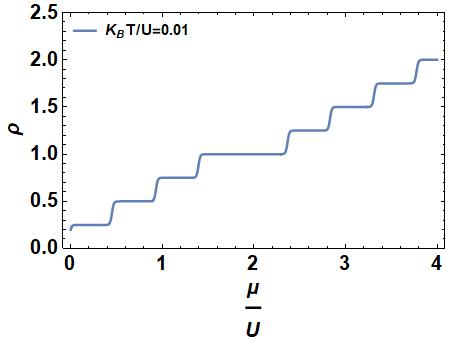}
    \caption{Number density plot}
    \label{fig:9}
\end{figure}

\begin{figure}
    \centering
    \includegraphics[width=0.9\linewidth]{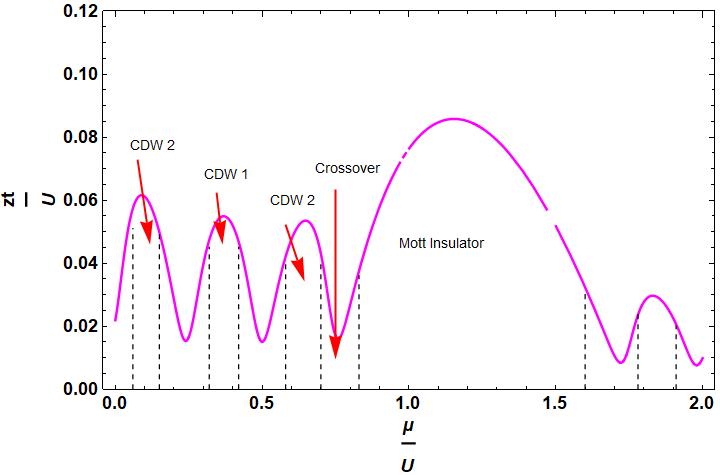}
    \caption{Phase diagram for $V=0.5U$ and $V^\prime=0.24U$ at temperature $K_BT=0.01U$ indicating different phases. They are MI, CDW 1, CDW 2, and normal fluid (cross-over region). The dashed line separates the normal fluid and insulator-like phases (MI, CDW1, and CDW 2).} 
    \label{fig:10}
\end{figure}

Now, for the finite temperature limit, we need to calculate  thermal averages of $(f_A+f_D)$ and $(f_B+f_C)$. For this, we first write the system's total unperturbed energy and partition function.
\begin{equation}
\begin{split}
    E^\prime(\mu)=&\frac{U}{2}n_A(n_A-1)+\frac{U}{2}n_B(n_B-1)+\frac{U}{2}n_C(n_C-1)\\
    &+\frac{U}{2}n_D(n_D-1)+\frac{zV}{2}(n_A+n_C)(n_B+n_D)\\
    &+zV^\prime(n_An_C+n_Bn_D)-\mu(n_A+n_B+n_C+n_D)
    \end{split}
\end{equation}
\begin{equation}
    Z^\prime=\sum_{n_A=0}^{\infty}\sum_{n_B=0}^{\infty}\sum_{n_C=0}^{\infty}\sum_{n_D=0}^{\infty}e^{-\beta{E^\prime}}
\end{equation}
Thermal averages of $(f_A+f_D)$ and $(f_B+f_C)$ will be,
\begin{equation*}
    \langle{f_A+f_D}\rangle=\frac{\sum_{n_A=0}^{\infty}\sum_{n_B=0}^{\infty}\sum_{n_C=0}^{\infty}\sum_{n_D=0}^{\infty}(f_A+f_D)e^{-\beta{E^\prime}}}{Z^\prime}
\end{equation*}
\begin{equation*}
\langle{f_B+f_C}\rangle=\frac{\sum_{n_A=0}^{\infty}\sum_{n_B=0}^{\infty}\sum_{n_C=0}^{\infty}\sum_{n_D=0}^{\infty}(f_B+f_C)e^{-\beta{E^\prime}}}{Z^\prime}
\end{equation*}
So, the phase  equation for non-zero temperature is
\begin{equation}\label{NNN3}
    \left\langle\frac{1}{zt}\right\rangle=\frac{1}{\sqrt{\langle{f_A+f_D}\rangle\langle{f_B+f_C}\rangle}}
\end{equation}
Eq.\eqref{NNN3} is plotted in Fig.\ref{fig:8} for different temperatures. In Fig.\ref{fig:8} the values of $V$ and $V^\prime$ (calculated from Rydberg potential curve in Sec. \ref{sec:Ryd}) are $0.96U$ and $0.48U$ respectively. Like the NN case, here a cross-over region with finite compressibility (Fig.\ref{fig:10}) also appears between the insulators. The dashed lines in Fig.\ref{fig:10} are drawn as the boundary of incompressible phases, i.e., at constant number density from Fig.\ref{fig:9}. With increasing temperature, the widths of the compressible lobes increase. As discussed earlier, the regions with compressibility $\kappa\leq0.01U$ are taken as incompressible regions. Like the NN interaction system, we obtain the transition temperature values up to which the incompressible phases survive. For the MI, it is the same as that obtained for NN in sec. \ref{pure} ($K_BT^*\thickapprox0.1U$), and for the two CDW phases it is approximately $0.08V$ (CDW $1$) and $0.04V^\prime$ (CDW $2$).


\subsection{Disordered system}\label{NNI+dis}
\begin{figure*}
    \centering
    \begin{subfigure}[b]{0.48\linewidth}
        \centering
        \includegraphics[width=0.9\textwidth]{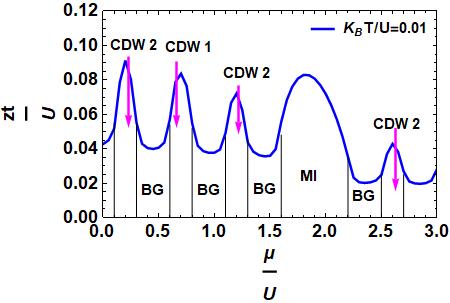}
        \caption{In presence of disorder with strength $\Delta=0.3U$, $V=0.9U$, $V^\prime=0.5U$. Interaction strengths are greater than $\Delta$. So all incompressible phases (MI, CDW 1, CDW 2) are present with BG between each of them. The vertical lines are drawn from the compressibility curve (fig \ref{fig:12}).}
        \label{NNN 1}
    \end{subfigure}
    \begin{subfigure}[b]{0.48\linewidth}
        \centering
        \includegraphics[width=0.9\textwidth]{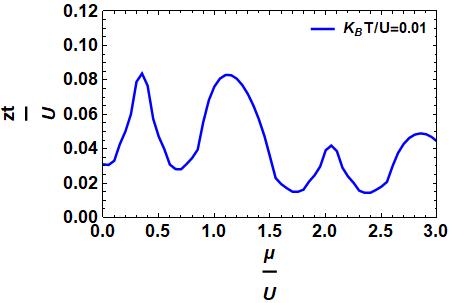}
        \caption{Disorder strength $\Delta=0.3U$, $V=0.6U$, $V^\prime=0.1U$, i.e. $\Delta$ is less than $V$ (NN), and greater than $V^\prime$ (NNN). So CDW 1 appears along with MI but the contribution of NNN is not there, hence CDW 2 is absent.}
        \label{NNN 2}
    \end{subfigure}
    \caption{Phase diagram for long-range interaction in the presence of disorder at different interaction potentials.}
    \label{fig:11}
\end{figure*}
\begin{figure}
    \centering
    \includegraphics[width=1.0\linewidth]{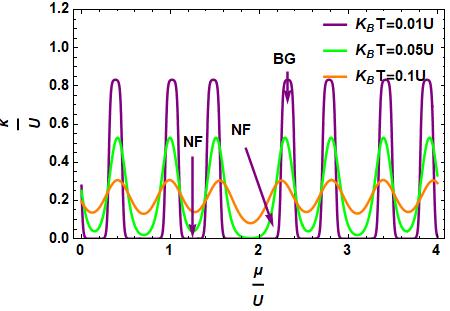}
    \caption{Compressibility indicating different phases at different temperatures at $\Delta=0.3U$. The arrows are for $K_BT=0.05U$ (green curves). It shows the presence of NF at finite temperature.}
    \label{fig:12}
\end{figure}
To add the disorder effect we do a disorder average of the two terms $\langle f_A+f_D\rangle$ and $\langle{f_B+f_C}\rangle$ in Eq. \eqref{NNN3} separately and combine them to get the phase diagram. Therefore, this phase diagram includes all the effects discussed in the earlier sections i.e. long-range interaction up to the next-nearest neighbor, non-zero temperature, and a uniform bounded disorder induced in the on-site potential. Fig. \ref{fig:11} shows the phase diagrams for different values of interaction strengths ($V$ and $V^\prime$) at a fixed temperature $K_BT=0.01U$ and disorder $\Delta=0.3U$. The phase diagram consists of three incompressible phases, the Mott Insulator, CDW 1, and CDW 2, and due to the presence of impurity a compressible Boss glass of width $\Delta$ between any two consecutive incompressible lobes. In Fig.\ref{NNN 1}, since $\Delta$ is less than both the interaction terms, all three incompressible phases along with BG are present. But in Fig.\ref{NNN 2} $\Delta$ is greater than NNN term $V^\prime$, so the CDW $2$ region is captured by BG.

 Now, to separate the incompressible phases from the compressible BG, we need to plot compressibility as a function of chemical potential. Detailed calculations are done in Appendix \ref{B}. 
The average density and compressibility can be written as,
\begin{equation}\label{density}
    \langle{\rho}\rangle=\frac{1}{2\beta\Delta}\ln\left[{\frac{\sum e^{-\beta E^\prime(\mu+\frac{\Delta}{2})}}{\sum e^{-\beta E^\prime (\mu-\frac{\Delta}{2})}}}\right]
\end{equation}
here, $\rho=(n_A+n_B+n_C+n_D)/4$ and sum is over $n_A$, $n_B$, $n_C$, $n_D$.
\begin{equation}\label{com3}
\begin{split}
    \kappa=&\frac{1}{4\Delta}\left[{\frac{\sum (n_A+n_B+n_C+n_D) e^{-\beta E^\prime\left(\mu+\frac{\Delta}{2}\right)}}{\sum  e^{-\beta E^\prime\left(\mu+\frac{\Delta}{2}\right)}}}\right.\\
    &\left.-{\frac{\sum (n_A+n_B+n_C+n_D) e^{-\beta E^\prime\left(\mu-\frac{\Delta}{2}\right)}}{\sum  e^{-\beta E^\prime\left(\mu-\frac{\Delta}{2}\right)}}}\right]
    \end{split}
\end{equation}
\begin{figure}
    \centering
    \includegraphics[width=1\linewidth]{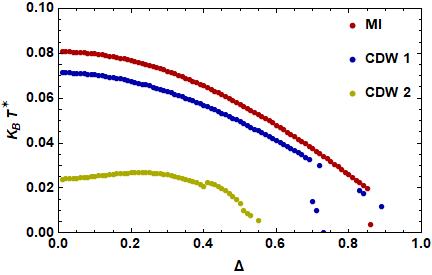}
    \caption{Change in the transition temperature ($K_BT^*$) with disorder strength $\Delta$ varying from $0$ to $1$ with $U$, $V$ and $V^\prime$ are fixed at $1$, $0.9$ and $0.5$ respectively for the three incompressible phases.} 
    \label{CriT 2}
\end{figure}
Fig.\ref{fig:12} shows the compressibility at different temperatures at $\Delta=0.3U$. At $K_BT=0.01U$ (purple curve), it is zero in the incompressible regions and finite for BG. The vertical lines in Fig.\ref{fig:11} are taken from here. As the temperature increases, the width of the compressible region increases, normal fluid emerges in the crossover. At $K_BT=0.05U$, $\kappa$ is finite in the CDW regions, that is, this region is completely melted into normal fluid (green curve). Above $K_BT=0.07U$, the Mott phase also vanishes, leaving only NF and BG (orange curve) to exist. Here, the transition or melting temperature for MI and CDW phases is a function of disorder and all interactions ($U$, $V$, $V^\prime$). Fig. \ref{CriT 2} shows the dependence of transition temperatures on the strength of the disorder $\Delta$, keeping the values of the interaction strengths fixed at $U=1$, $V=0.9$, $V^\prime=0.5$. It is clear from Fig.\ref{CriT 2} that $T^*$ decreases with increasing disorder and falls to zero when disorder is equal to the corresponding interaction strength.

\section{Conclusions}\label{con}
In this paper, we discuss the effect of temperature on different phases and phase transitions in the extended Bose-Hubbard model in both the absence and presence of disorder. We take a Rydberg-atom chain as a prototype of such a model, and illustrate how (i) nearest-neighbor (NN) interaction only, and (ii) interactions up to the next-nearest-neighbor(NNN) can be implemented using such a Rydberg lattice: by adjusting the lattice constant with respect to the Rydberg blockade radius.  As for the disorder, the form that we consider is a uniform and bounded disorder. 

In the pure system, when the interaction is up to the nearest neighbor (Sec.\ref{pure}), a compressible normal fluid emanates between the incompressible CDW and MI phase at non-zero temperatures. It is found that with increasing temperature, this normal phase extends in height and width in the $zt$ vs. $\mu$ phase diagram, causing the superfluid phase to shift towards a higher value of the hopping parameter. This is accompanied by a contraction of the area of the incompressible lobes. Above a certain temperature, the CDW lobes are destroyed and a normal fluid replaces them. The temperature corresponding to the Mott lobes is higher, but eventually, they, too, melt into a normal fluid. 

When both NN and NNN interactions are present, two types of CDW phases appear at zero temperature. One is of density ${n}/{4}$ (CDW 2, arising from NNNI $V^\prime$) and the other is of $n/2$ (CDW 1, arising from NNI $V$), where $n=1,3,5,.....$. The zero temperature diagram is identical to that observed in \cite{kovrizhin2005density} by the Gutzwiller variational approach. We observe that at non-zero temperature, these lobes suffer the same fate, i.e, they melt and become normal fluid. In this case, the temperature for CDW 2 is the lowest, followed by CDW 1 and MI, respectively.

In a similar way, interactions up to several sites can be introduced by choosing a suitable combination of lattice spacing and Rydberg blockade. In that case, more CDW phases will appear with different density structures. However, since the Rydberg potential curve falls off rapidly (Fig:\ref{ryd2}), $V'' << V'$, and  $V'''<< V''$, and so on - if $V''$ and $V'''$ are the next order of interaction terms. Therefore, the width of the new CDW phases emerging due to longer-range interactions is very small. Not only the lobe-width, but also the temperature is proportional to these interaction strengths.  Thus, these lobes survive only for very small temperatures.

If the system is embedded with a uniform disorder, the Boss glass phase emerges between the incompressible lobes. At finite temperature, normal fluid comes in the crossover of BG and incompressible lobes. With increasing temperature, MI and CDW melt into NF, but BG exists even at high temperature as a signature of disorder. In the presence of disorder, the transition temperature depends not only on the interaction strengths  $U$, $V$, and $V^\prime$, but also on the amount of disorder. It is shown that at fixed values of the interaction strength, $T^*$ decreases with increasing $\Delta$. 

The mathematical framework that we present here is capable of handling long-range interaction, disorder, and non-zero temperature all at the same time. By taking appropriate limits, one can recover the results of (i) zero-temperature EBHM, (ii) a finite-temperature disordered system with on-site interaction only, and (iii) a zero-temperature system with nearest-neighbor interaction and disorder. Moreover, the generic framework is versatile enough so that even longer-range (i.e., beyond the NNN) interactions and different forms of disorder can be incorporated here easily.

In future, we would like to use this framework to include other possible forms of disorder at non-zero finite temperatures. A study of some other statistical measures like entanglement entropy is also on the cards. A possible interesting direction would be to incorporate the effect of fluctuations on top of the mean-field results, and investigate how it affects the melting temperatures of the insulating lobes.

\section{Acknowledgements:} The authors thank Indrakshi Raychowdhury for fruitful discussions. RD would like to acknowledge Anusandhan National Research Foundation (ANRF) (erstwhile Science and Engineering Research Board (SERB)), Department of Science and Technology, Govt.of India for providing support under the CRG scheme (CRG/2022/007312).
MK would like to acknowledge Council of Scientific and Indrustrial Research (CSIR), Govt. of India, for financial support \big(file no.- 09/0028(21026)/2025-EMR-I\big).

\appendix
\section{Appendix}
   
\subsection{Order parameter calculation for long range (NN and NNN) interaction}\label{A}
For one sublattice (say \textbf{A}) the unperturbed energy is,
\begin{equation}
    H_{Un}=\frac{U}{2}n_A(n_A-1)+\frac{zV}{2}n_A(n_B+n_D)+zV^\prime{n_A}n_C-\mu{n_A}
\end{equation}
and the perturbation is,
\begin{equation}
    H^\prime=-\frac{zt}{2}(\psi_B+\psi_D)(b_A+b_A^\dagger)+zt\psi_A(\psi_B+\psi_D)
\end{equation}

So, the SF order parameter for sublattice \textbf{A} is given by,
\begin{equation*}
    \psi_{A} = \langle\chi_{A}|\hat{b_{A}}|\chi_{A}\rangle
\end{equation*}
 	
\begin{equation}
\begin{split}
	\label{eqB1}
	\implies \psi_{A} = & \langle{n_{A}}|\hat{b_{A}}|{n_{A}}\rangle + \sum\limits_{m_{A}\neq{n_{A}}}\frac{\langle{m_{A}|\hat{H}^{\prime}|n_{A}\rangle}}{{E_n}_{A} - {E_m}_{A}}\,\langle{m_{A}}|\hat{b}_{A}|n_{A}\rangle 
	\\& + \sum\limits_{m_{A}\neq{n_{A}}}\frac{\langle{m_{A}|\hat{H}^{\prime}|n_{A}\rangle}}{{E_n}_{A} - {E_m}_{A}}\,\langle{n_{A}}|\hat{b}_{A}|m_{A}\rangle \\
 &+
\sum\limits_{m_{A}\neq{n_{A}}}\left(\frac{\langle{m_{A}|\hat{H}^{\prime}|n_{A}\rangle}}{{E_n}_{A} - {E_m}_{A}}\right)^2\,\langle{m_{A}}|\hat{b_{A}}|m_{A}\rangle
\end{split}
\end{equation}
The first and last terms are zero. We assume that all $\psi$'s are real i.e. $\psi_{A} = \psi_{A}^*$.
\begin{equation*}
\langle{m_{A}}|\hat{H}^{\prime}|n_{A}\rangle  = \langle{m_{A}}|-\frac{zt}{2}(\psi_B+\psi_D)(b_A+b_A^\dagger)+zt\psi_A(\psi_B+\psi_D)
    |n_{A}\rangle
\end{equation*}
(The second term is zero because it contributes only for $m_{A} = n_{A}$ which is not acceptable.) Hence,

\begin{equation*}
    \begin{split}
        \langle{m_{A}}|\hat{H}^{\prime}|n_{A}\rangle&=-\frac{zt}{2}(\psi_B+\psi_D)\times\\
        &\left[\sqrt{n_A+1}\delta_{m_A,n_A+1}
        +\sqrt{n_A}\delta_{m_A,n_A-1}\right]
    \end{split}
\end{equation*}

Now from Eq. \ref{eqB1} we have,
\begin{align}\label{B2}
E_{n_{A}} - E_{n_{{A}-1}} & = U(n_{A} - 1) - \mu + \frac{zV}{2}(n_B+n_D)+zV^{\prime}n_C\\
\label{B3}
E_{n_{A}} - E_{n_{{A}+1}} & = \mu -Un_{A} - \frac{zV}{2}(n_B+n_D)-zV^{\prime}n_C
\end{align}
Substituting all these in Eq. \ref{eqB1} we finally get,
\begin{equation}
\psi_{A}=\frac{zt}{2}(\psi_B+\psi_D)\times{f_A}
\label{A6}
\end{equation}
where,\begin{equation}
        \begin{split}
         f_A=&\left[\frac{n_{A} + 1}{Un_{A} + \frac{zV}{2}(n_B+n_D)+zV^{\prime}n_C -\mu}\right.
     \\
     &-\left.\frac{n_{A}}{U(n_{A} - 1) + \frac{zV}{2}(n_B+n_D)+zV^{\prime}n_C - \mu}\right] 
        \end{split}
\end{equation}

 Similarly, for the other three sublattices, the order parameters are,
 \begin{equation}
\begin{split}
    &\psi_B=\frac{zt}{2}\left(\psi_A+\psi_C\right)\times{f_B}\\
    &\psi_C=\frac{zt}{2}\left(\psi_B+\psi_D\right)\times{f_C}\\
    &\psi_D=\frac{zt}{2}\left(\psi_A+\psi_C\right)\times{f_D}
    \end{split}
    \label{A8}
\end{equation}
where,
\begin{equation*}
        \begin{split}
         f_B=&\left[\frac{n_{B} + 1}{Un_{B} + \frac{zV}{2}(n_A+n_C)+zV^{\prime}n_D -\mu}\right.
     \\
     &-\left.\frac{n_{B}}{U(n_{B} - 1) + \frac{zV}{2}(n_A+n_C)+zV^{\prime}n_D - \mu}\right] 
        \end{split}
\end{equation*}
 \begin{equation*}
        \begin{split}
         f_C=&\left[\frac{n_{C} + 1}{Un_{C} + \frac{zV}{2}(n_B+n_D)+zV^{\prime}n_A -\mu}\right.
     \\
     &-\left.\frac{n_{C}}{U(n_{C} - 1) + \frac{zV}{2}(n_B+n_D)+zV^{\prime}n_A - \mu}\right] 
        \end{split}
\end{equation*}
\begin{equation*}
        \begin{split}
         f_D=&\left[\frac{n_{D} + 1}{Un_{D} + \frac{zV}{2}(n_A+n_C)+zV^{\prime}n_B -\mu}\right.
     \\
     &-\left.\frac{n_{D}}{U(n_{D} - 1) + \frac{zV}{2}(n_A+n_C)+zV^{\prime}n_B - \mu}\right] 
        \end{split}
\end{equation*}
 
\subsection{Disordered average of compressibility of the insulating phases for extended model}\label{B}
In \cite{bissbort2010stochastic} they calculated average particle density and compressibility for the disordered system. We use the same formulation in our extended case.
The disordered average of boson density for a system with total energy $E$ will be,
\begin{equation}
\begin{split}
         \langle{\rho}\rangle &= \int\frac{\sum e^{-\beta E(\mu+\epsilon)}\times\rho}{Z}p(\epsilon)d\epsilon\\
         &=\frac{1}{\beta}\int\frac{\partial\rho(\epsilon)}{\partial\epsilon}\left(\ln{\sum e^{-\beta E(\mu+\epsilon)}}\right)
         \label{den3}
         \end{split}
    \end{equation}
    where $E(\mu+\epsilon)$ is the system's total unperturbed energy. For NN interaction it is, $E_{n_A,n_B}(\mu+\epsilon)=({U}/{2})n_A(n_A-1)+({U}/{2})n_B(n_B-1)+zVn_An_B-(\mu+\epsilon)(n_A+n_B)$. $\rho=(n_A+n_B)/{2}$ and the sum is over $n_A$ and $n_B$. 
    
Therefore, for uniform disorder, the average takes the following form,
\begin{equation}\label{den2}
    \langle{\rho}\rangle=\frac{1}{2\beta\Delta}\ln\left[{\frac{\sum_{n_A=0}^\infty\sum_{n_B=0}^\infty e^{-\beta E_{n_A,n_B}(\mu+\frac{\Delta}{2})}}{\sum_{n_A=0}^\infty\sum_{n_B=0}^\infty e^{-\beta E_{n_A,n_B}(\mu-\frac{\Delta}{2})}}}\right]
\end{equation}
So, the compressibility $\kappa=\frac{\partial{\langle{\rho}\rangle}}{\partial\mu}$ for the disordered EBHM with NN  will be,
\begin{equation}\label{com1}
\begin{split}
    \kappa=&\frac{1}{2\Delta}\left[{\frac{\sum_{n_A=0}^\infty\sum_{n_B=0}^\infty (n_A+n_B) e^{-\beta E_{n_A,n_B}\left(\mu+\frac{\Delta}{2}\right)}}{\sum_{n_A=0}^\infty\sum_{n_B=0}^\infty  e^{-\beta E_{n_A,n_B}\left(\mu+\frac{\Delta}{2}\right)}}}\right.\\
    &\left.-{\frac{\sum_{n_A=0}^\infty\sum_{n_B=0}^\infty (n_A+n_B) e^{-\beta E_{n_A,n_B}\left(\mu-\frac{\Delta}{2}\right)}}{\sum_{n_A=0}^\infty\sum_{n_B=0}^\infty  e^{-\beta E_{n_A,n_B}\left(\mu-\frac{\Delta}{2}\right)}}}\right]
    \end{split}
\end{equation}
Similarly, for the next-nearest neighbor interaction, the compressibility will be,
\begin{equation}\label{com2}
\begin{split}
    \kappa=&\frac{1}{4\Delta}\left[{\frac{\sum_{n_A=0}^\infty\sum_{n_B=0}^\infty \sum_{n_C=0}^\infty\sum_{n_D=0}^\infty(n_A+n_B+n_C+n_D) e^{-\beta E^{\prime}\left(\mu+\frac{\Delta}{2}\right)}}{\sum_{n_A=0}^\infty\sum_{n_B=0}^\infty \sum_{n_C=0}^\infty\sum_{n_D=0}^\infty  e^{-\beta E^{\prime}\left(\mu+\frac{\Delta}{2}\right)}}}\right.\\
    &\left.-{\frac{\sum_{n_A=0}^\infty\sum_{n_B=0}^\infty \sum_{n_C=0}^\infty\sum_{n_D=0}^\infty(n_A+n_B+n_C+n_D) e^{-\beta E^{\prime}\left(\mu-\frac{\Delta}{2}\right)}}{\sum_{n_A=0}^\infty\sum_{n_B=0}^\infty \sum_{n_C=0}^\infty\sum_{n_D=0}^\infty  e^{-\beta E^{\prime}\left(\mu-\frac{\Delta}{2}\right)}}}\right]
    \end{split}
\end{equation}

where $E^{\prime}$ is the total unperturbed energy for the NNN system.

\bibliography{references}

\end{document}